\documentclass[prc,twocolumn,floatfix,groupedaddress,nofootinbib,preprintnumbers,amsmath,amssymb,amsfonts,superscriptaddress,widetable,href]{revtex4}
\usepackage{amsmath,amssymb,amsfonts}
\usepackage{graphicx}
\usepackage{color}
\usepackage{hyperref}
\usepackage{ulem}

\allowdisplaybreaks

\newcommand{\be}{\begin{equation}}
\newcommand{\ee}{\end{equation}}
\newcommand{\bea}{\begin{eqnarray}}
\newcommand{\eea}{\end{eqnarray}}

\newcommand{\al}{\alpha}

\newcommand{\de}{\delta}
\newcommand{\De}{\Delta}

\newcommand{\Th}{\Theta}

\begin{document}

\setcounter{page}{1}

\vspace*{0.5 true in}

\title{The nucleon effective mass and its isovector splitting}
\author{Tuhin \surname{Malik}}
\email{tuhin.malik@gmail.com}
\affiliation{Department of Physics, BITS-Pilani, K. K. Birla Goa Campus,
Goa 403726, India}
\author{C. \surname{Mondal}}
\email{chiranjib.mondal@icc.ub.edu}
\affiliation{Saha Institute of Nuclear Physics, 1/AF 
 Bidhannagar, Kolkata 700064, India.}  
 \affiliation{Departament de F\'isica Qu\`antica i Astrof\'isica and Institut de
Ci\`encies del Cosmos (ICCUB), Facultat de F\'isica, Universitat de Barcelona, Mart\'i
i Franqu\`es 1, E-08028 Barcelona, Spain}
\author{B. K. \surname{Agrawal}}
\email{bijay.agrawal@saha.ac.in}
\affiliation{Saha Institute of Nuclear Physics, 1/AF 
 Bidhannagar, Kolkata 700064, India.}  
\affiliation{Homi Bhabha National Institute, Anushakti Nagar, Mumbai 400094, India.}  
\author{J.N. \surname{De}}
\email{jn.de@saha.ac.in}
\affiliation{Saha Institute of Nuclear Physics, 1/AF 
 Bidhannagar, Kolkata 700064, India.}  
\author{S.K. \surname{Samaddar}}
\email{santosh.samaddar@saha.ac.in}
\affiliation{Saha Institute of Nuclear Physics, 1/AF 
 Bidhannagar, Kolkata 700064, India.}  

\begin{abstract} 
 Using an energy density functional (EDF) based on
the thermodynamic Gibbs-Duhem relation, found equivalent to the
standard Skyrme EDF for infinite nuclear matter, it is  demonstrated
that the parameters of this EDF are not uniquely determined from the
fit of the empirical and theoretical data related to nuclear matter.
This prevents an unambiguous determination of the nucleon effective
mass ($\frac{m_0^*}{m}$) and its isovector splitting ($\De m_0^*$).
Complementary information from dipole polarizability of atomic nuclei
helps in removing this ambiguity and plausible values of $\frac{m_0^*}{m}$
and $\Delta m_0^*$ can be arrived at. Presently considered fit data
on infinite nuclear matter and dipole polarizability of finite nuclei
yield $\frac{m_0^*}{m} =0.68 \pm 0.04 $ and $\De m_0^* =(-0.20 \pm
0.09)\delta $.  This  EDF is consistent with the constraint  on the
maximum mass of the neutron star.

\end{abstract}



\maketitle

\section{Introduction}

The nucleon effective mass is a measure of the mobility of the nucleon
in a nuclear medium. It is defined as $d\epsilon/dp=p/m^* $, where $m^*$ is the
effective nucleon mass, $\epsilon $ the energy per nucleon in the system and
$p$ is the magnitude of the nucleon momentum. For a homogeneous system
of density $\rho$, $m^*(\rho )$ is defined at the Fermi surface.  In 
isospin asymmetric nuclear matter, the neutrons and protons may feel 
different single-particle potentials. This may result in a difference in
their mobility. The (neutron, proton) effective masses ($m_n^*, m_p^* $)
may then be different. The isospin-splitted effective mass is defined as 
$\De m^*=(m_n^*-m_p^*)/m $ where $m$ is the bare mass of the nucleon. The
small difference between the bare neutron and the bare proton mass is neglected.
There exists different kinds of nucleon effective masses in relativistic
\cite{Serot86,Chen07} and non-relativistic approaches \cite{Mahaux85,Jaminon89},
we confine ourselves specifically to a non-relativistic description. Furthermore,
only that part of the mass renormalization coming from the momentum dependence
of the nucleon-nucleon interaction is in focus here. The energy-mass component
emanating from the coupling of the nucleon effective mass with the dynamical
vibration \cite{Bohr75,Hasse86} of the single-particle potential is left
out.

Attempts have been made in the last few decades to arrive at an acceptable 
value for the nucleon effective mass $m^*(\rho_0) (=m_0^*)$ at the saturation
density $\rho_0$ of symmetric nuclear matter (SNM), but a clear consensus 
still seems to be missing. Energy density functionals (EDF), many in the 
Skyrme framework, designed to effectively reproduce various empirical 
properties of nuclear matter and finite nuclei yield values of $m_0^*/m$
in the range 0.6-1.0 \cite{Brown80,Chen09,Dutra12,Davesne18}. 
Many-body calculations, 
irrespective of their level of sophistication give $m_0^*/m \sim 0.8
\pm 0.1$ \cite{Friedman81,Wiringa88,Zuo99}. Analysis of isoscalar 
giant quadrupole resonances (ISGQR) \cite{Zhang16,Bender03,Stone07,Roca-Maza13}
predicts a similar value ($\sim 0.85\pm 0.1$), but the analysis 
is not model independent.
Optical model analysis of nucleon-nucleus scattering, on the other hand,
yield a value of the effective mass somewhat less, $m_0^*/m \sim 0.65 
\pm 0.06$ \cite{Li15b}. 

The isovector sector of the nuclear interaction is mired with large 
uncertainty. The symmetry energy coefficient, its density
derivatives of different orders and  the isovector splitting of the nucleon 
effective mass offer a window to have a close glimpse on the nature
of this part of the interaction. The isovector mass splitting
is of profound importance in addressing many
key problems in nuclear physics, astrophysics and even cosmology.
It is critical for understanding transport and thermal properties of
asymmetric nuclear matter \cite{Xu15,Behera11,Xu15a,Li04}, for 
neutrino opacities in neutron star matter \cite{Burrows06,Baldo14},
for locating the  drip lines \cite{Woods97} in the nuclear mass table.
It has also a crucial role to play in understanding neutron/proton ratio
in primordial nucleosynthesis \cite{Steigman06} in the early universe.

Attempts have been made in recent times to have an experimental 
estimation of the isovector effective mass splitting $\De m_0^*$
(the subscript refers to its value at $\rho_0$). From analysis of
nucleon-nucleus scattering data \cite{Li15b} within an isospin-dependent
optical model, it is found to be $(0.41 \pm 0.15)\de$ ($\de $ is the
isospin asymmetry defined as $(\rho_n-\rho_p)/\rho) $, from exploration
of ISGQR and dipole polarizability \cite{Zhang16} it goes down to
$(0.27 \pm 0.15)\de $. On the other hand, the transport-model-motivated
estimates of double $n/p $ ratio in heavy ion collisions suggest
\cite{Kong17} a value of $\De m_0^*=(-0.058 \pm 0.129)\de $. 
Theoretical studies based on microscopic or phenomenological 
approaches have also yielded varying values of $\De m_0^* $ \cite{Zuo05,
Dalen05,Bandyopadhyay90,Mondal17}. Non-relativistic Brueckner-
Hartree-Fock (BHF) and relativistic Dirac-Brueckner-Hartree-Fock (DBHF)
calculations \cite{Zuo05,Dalen05} give $\De m_0^* > 0$ whereas
Gogny-Hartree-Fock models \cite{Baran05,Chen12,Sellahewa14} result in 
$\De m_0^* $ positive or negative depending on 
the choice of the parameters defining the force.
On the phenomenological side, for around 100 Skyrme EDFs , $\De m_0^*$
comes out to be positive for about half of them, around one-third
are negative and the rest are nearly zero \cite{Ou11}. From constraints on
Skyrme EDFs provided by properties of nuclear matter \cite{Dutra12},
of doubly magic nuclei and ab-initio calculations of low-density
neutron matter, recently some "best-fit" EDFs were
isolated  and listed in Table I of Ref. \cite{Brown13}.
From the parameters of these EDFs (with the values of $x_1$ as
given in the last column of the Table), 
the isovector-splitted
effective mass can be calculated. For all of them, $\De m_0^*$ is
found to be negative ($\simeq -0.3\de $). Both experiment and theory,
however, thus far seem to weigh  in favor of a positive 
$\De m_0^*$. In this conundrum, a fresher look at  the nucleon 
effective mass, its isovector splitting and thus a closer view of
the isovector sector of the nuclear interaction is called for. 
This article is an attempt towards this goal, finding a 
means to determine the values of $\frac{m_0^*}{m}$ and $\Delta m_0^*$ 
in a subtle way.

  The paper is organized as follows. In Sec. II, the elements comprising 
the theoretical background are reviewed. The results and discussions are
presented in Sec. III. The concluding remarks are drawn in Sec. IV.

\section{Theoretical framework}

In an effort to find the interdependence of the different symmetry
energy elements of nuclear matter, a specific EDF was recently
constructed \cite{Mondal17} structured on the thermodynamic Gibbs-Duhem
relation. Built in the confines of non-relativistic mean field approximation,
no specific assumption about the nuclear interaction is made in this
EDF except that it is effectively two-body, 
quadratically momentum dependent and that
it has a power law density dependence to simulate many-body effects. 
Equations relevant in the present context are
presented in subsection A, subsection B contains a discussion
on nuclear dipole polarizability and its relation in finding
some key parameters of this EDF uniquely. 

\subsection{The energy density functional}
Exploiting Gibbs-Duhem relation,
the energy per nucleon of asymmetric nuclear matter at a density
$\rho $ and isospin asymmetry $\delta $ can be written as \cite{Mondal17} 
\bea
\label{ener3}
\epsilon(\rho, \de)&=&\frac{1}{\rho} \left[\sum_\tau \frac{P_{F,\tau}^2}{10m}
\rho_\tau 
\left(3-2\frac{m}{m^*_\tau (\rho)}\right)  \right]-V_2(\rho,\de) \nonumber \\
&&+\frac{P(\rho,\de) }{\rho}.
\eea 
The index $\tau $ refers to neutron or proton, 
$\rho_\tau =(1+\tau \de)\rho /2$;
$\tau $=1 for neutrons and -1 for protons. The Fermi momentum
$P_{F,\tau} =g\rho_\tau^{1/3}$ with $g=(3\pi^2)^{1/3}\hbar$. 
In Eq. (\ref{ener3}), $P(\rho,\delta )$ is the pressure of the system 
and $V_2(\rho,\delta )$ is the rearrangement potential. 
The density and isospin dependent rearrangement potential can be expanded
around $\de =0$ and written as 
\bea
\label{sp2}
V_2(\rho,\de)=(a+b\de^2+c\de^4 +\dots ) \rho^{\tilde \al}.
\eea
If the interaction is effectively  two-body, terms beyond
$\delta^2$ are zero.
The density-dependent nucleon effective mass is taken as 
\bea
\label{mstar2}
\frac{m}{m_\tau^*(\rho )}=1+\frac{k_+}{2}\rho +\frac{k_-}{2}\rho \tau \de.
\eea
The isovector effective mass splitting $\De m_0^* \left(=\frac{m_n^*-m_p^*}{m}\right) $
at $\rho_0$ is then given by 
\bea
\label{dmstar1}
\De m_0^*\simeq -k_-\rho_0 \left(\frac{m^*_0}{m}\right)^2 \de,
\eea
where at the saturation density, the approximation $(m_n^*.m_p^*) \simeq 
(m_0^*)^2$ is used.

Since the pressure $P=\rho^2\frac{\partial \epsilon}{\partial \rho}$, Eq.
(\ref{ener3}) can be integrated to
\bea
\label{ener4}
\epsilon(\rho,\de)&=&\frac{3}{2}y\Bigl [\sum_\tau (1+\tau \de)^{5/3}
\bigl \{ \rho^{2/3}+\frac{1}{2}\rho^{5/3}(k_++k_-\tau\de)\bigr \} \Bigr ]
\nonumber \\
&&+(a+b\de^2)\frac{\rho^{\tilde \al}}{(\tilde \al -1)}+K(\de)\rho,
\eea
where $y=\frac{g^2}{10m.2^{2/3}}$ and $K(\de)=(K_1+K_2\de^2 +K_4\de^4 +
\cdots )$ is a constant of integration.

The expressions for the energy of SNM and pure neutron matter (PNM) and their
pressures are then written as
\bea
\label{ed0}
\epsilon(\rho,\de =0)=a\frac{\rho^{\tilde \al}}{(\tilde \al -1)} +3y\rho^{2/3}
\left(1+\frac{1}{2}k_+\rho\right)+K_1\rho\ 
\eea
\bea
\label{ed1}
&\epsilon&(\rho,\de =1)=(a+b)\frac{\rho^{\tilde \al}}{(\tilde \al -1)}  
+3\times 2^{2/3}y\rho^{2/3} \times\nonumber \\ 
&&\left[1+\frac{1}{2}(k_+ +k_-)\rho\right]  
+(K_1+K_2+K_4+\cdots )\rho\ 
\eea
\bea
\label{pd0}
P(\rho,\de =0)=\frac{\tilde \al}{(\tilde \al -1)}a\rho^{\tilde \al +1}
\nonumber \\
+y\rho^{5/3}\left(2+\frac{5}{2}k_+\rho\right)+K_1\rho^2
\eea
\bea
\label{pd1}
&&P(\rho,\de =1)=\frac{\tilde \al}{(\tilde \al -1)}(a+b)\rho^{\tilde \al +1}
+2^{2/3}y\rho^{5/3} \times\nonumber \\
&&\left[2+\frac{5}{2}(k_++k_-)\rho\right]  
+(K_1+K_2+K_4+\cdots )\rho^2.
\eea
From Eq.(\ref{ener4}), the symmetry energy coefficient $C_2(\rho )
(=\frac{1}{2}\frac{\partial^2\epsilon }{\partial \de^2}|_{\de =0})$ is derived
as
\bea
\label{c21}
C_2(\rho)&=&\frac{b\rho^{\tilde \al}}{(\tilde \al -1)} +\frac{5}{3}y\rho^{2/3}
\nonumber \\
&&\times \left[1+\frac{1}{2}(k_+ +3k_-)\rho\right] +K_2\rho.
\eea

The parameters of this EDF can be found from the best fit to the existing
empirical data pertaining to nuclear matter, i.e., the pressure of 
symmetric nuclear matter \cite{Danielewicz02,Fuchs06,Fantina14} and 
of pure neutron matter (PNM) in a broad density range \cite{Danielewicz02,
Prakash88} and also the symmetry energy in a limited density range
\cite{Tsang09,Tsang10,Danielewicz14,Russotto16}. In addition, the energy and
pressure of low density neutron matter calculated in high precision in 
chiral effective field theory ($N^3LO$) \cite{Hebeler13,
Sammarruca15} are taken into account in the fit data. Our analysis
reveals that the fit to the infinite nuclear matter data alone 
is unable to fix the values of the two parameters 
($k_+$ and $k_-$) separately; it 
tends to yield a value of the sum of the parameters.  Available 
experimental data on the dipole polarizability in a few nuclei, on the
other hand, are shown to illuminate the relation of an
isovector property with the 
difference between these said parameters ($k_+$ and $k_-$).
In this paper, we use  this extra information to find values of 
$k_+$ and $k_-$; they are measures of the 
nucleon effective mass and the isovector mass splitting. 

Inspection of the EDF in Eq.(\ref{ener4}), when compared with the
'standard' Skyrme functional \cite{Dutra12} shows that there is an exact
equivalence of the Skyrme functional for infinite matter with the one
given by Eq. (\ref{ener4}) provided the term $K(\de )$ is truncated
at $\de^2$. In subsequent analysis, we take this prescription, i.e; $K_4$
and higher order terms are ignored.
The parameters $\tilde \al, K_1,K_2,a,b, k_+ $ and $k_-$ can
then be correlated to the standard Skyrme parameters:
\bea
\label{sk1}
\tilde \al&=&\al +1   \nonumber \\
K_1&=&\frac{3}{8}t_0  \nonumber \\
K_2&=&-\frac{1}{4}t_0\left(x_0+\frac{1}{2}\right) \nonumber \\
a&=&\frac{1}{16}t_3\al  \nonumber \\
b&=&-\frac{1}{24}t_3\left(x_3+\frac{1}{2}\right)\al \nonumber \\
k_+&=&\frac{m}{\hbar^2}\left[\frac{3}{4}t_1+\frac{5}{4}t_2+t_2x_2\right] \nonumber \\
k_-&=&\frac{m}{2\hbar^2}\left[t_2\left(x_2+\frac{1}{2}\right)-t_1\left(x_1+\frac{1}{2}\right)\right].
\eea

\subsection{The isovector mass, the  energy weighted sum rule 
and  dipole polarizability}

 Eq. (\ref{mstar2}) shows that the parameters $k_+$
and $k_-$ can be used to define the 
nucleon effective mass and the isovector mass
splitting. It also defines, in terms of these parameters, 
the isovector nucleon mass $m_{v,0}^*$,
i.e; the effective mass of a proton in pure neutron matter or vice versa
at $\rho_0$ \cite{Zhang16}.
It is given as
\bea
\label{mv1}
\frac{m}{m_{v,0}^*}=1+\frac{m}{2\hbar^2}\rho_0 \Theta_V
\eea
where 
\bea
\label{tv1}
\Theta_V=\frac{\hbar^2}{m}(k_+-k_-).
\eea
An added knowledge of $\Theta_V$ helps in finding $k_+$ and $k_-$.
In Skyrme parameterization, the isovector parameter $\Theta_V$ is
\bea
\label{tv2}
\Theta_V=\left[t_1\left(1+\frac{x_1}{2}\right)+t_2\left(1+\frac{x_2}{2}\right)\right].
\eea

The  parameter $\Theta_V$ is related  to $m_1$, the energy
weighted sum rule (EWSR) for the dipole excitations. The $k$-th moment
of the energy weighted sum is defined as 
\bea
\label{mk}
m_k=\int dE E^k S(E) ,
\eea
where $S(E)$ is the strength function at energy $E$. For the isovector 
giant dipole resonance (IVGDR) of a nucleus with mass number $A$,
neutron number $N$ and proton number $Z$, the EWSR can be written as
\cite{Bohr75},
\bea
\label{mp1}
m_1=\frac{9}{4\pi}\frac{\hbar^2}{2m}\frac{NZ}{A}(1+\kappa_A),
\eea
where $\kappa_A$ is the polarizability enhancement factor for the
nucleus in question. It is related to $\Theta_V$ as \cite{Chabanat97} 
\bea
\label{kap1}
\kappa_A=\frac{2m}{\hbar^2}\frac{A}{4NZ}\Theta_V \times I_A 
\eea
where $I_A= \int \rho_n(r)
\rho_p(r)d^3r $; $\rho_n(r)$ and $\rho_p(r)$ are the neutron and 
proton density distributions of the nucleus.  In principle, $m_1$
can be found out from the experimentally determined strength function
$S(E)$; it is then possible to get to a value of $\Theta_V$ provided
the integral occurring in Eq.(\ref{kap1}) is known.
However, the high energy component of the strength function is plagued
with 'quasi-deuteron effect' rendering the determination of 
$m_1$ or $\kappa_A$ not very reliable \cite{Lepretre81, Schelhass88}.

 It need be mentioned that experimental data for the inverse
energy weighted sum $m_{-1}$ for a few nuclei \cite{Birkhan17,Rossi13,
Hashimoto15,Tamii11}, or in other words, the nuclear dipole polarizability  
are available.  They are related as
\bea
\label{ald1}
\al_D=\frac{8\pi e^2}{9}\int dE E^{-1}S(E)=\frac{8\pi e^2}{9}m_{-1}.
\eea
Using Eqs. (\ref{mp1}) and (\ref{ald1}), one can then  write
\bea
\label{ald2}
m_1=\frac{9}{8\pi e^2}E_x^2\al_D,
\eea
where the energy $E_x=(\frac{m_1}{m_{-1}})^{1/2}$ is referred to as the
constrained energy \cite{Bohigas79}.  
To find $\Th_V$, values of $m_1$ are constructed from reasonable inputs
on $E_x$ which we discuss in the next section.

\section{Results and  discussions}
\begin{table}
  \caption{List of fit data ($P(\rho)$, ${\epsilon}_n(\rho)$ and $C_2(\rho)$ represent pressure, energy per particle and symmetry energy 
     respectively) corresponding to the symmetric nuclear matter (SNMX), pure neutron matter (PNMX) and symmetry energy coefficient 
     (SYMX) together with the range of densities in which they are
determined.}
  \label{tab1}
      \setlength{\tabcolsep}{3.5pt}
      \renewcommand{\arraystretch}{1.1}
     \begin{tabular}{ l c c c c }
      \toprule
     
           & Quantity & Density region & Band/Range & Ref.  \\ 
            &     & $(\text{fm}^{-3})$ & (MeV) & \\
      \hline 
      
      SNM1 & $P(\rho)$ 	   &  $0.32~\text{to}~0.74$   &  HIC        & \cite{Danielewicz02}  \\
      SNM2 & $P(\rho)$     &  $0.19~\text{to}~0.33$   &  Kaon exp       & \cite{Fuchs06,Fantina14}  \\
      \\
      PNM1 & ${\epsilon}_{n}(\rho)$  & 0.1            &   $10.9\pm0.5$     & \cite{Brown13} \\
      PNM2 & ${\epsilon}_{n}(\rho)$   & $0.03~\text{to}~0.17$   &   N$^{3}$LO        & \cite{Hebeler13} \\
      PNM3 & $P(\rho)$            & $0.32~\text{to}~0.73$   &   HIC         & \cite{Danielewicz02} \\
      PNM4 & $P(\rho)$            & $0.03~\text{to}~0.17$   &   N$^{3}$LO        & \cite{Hebeler13} \\
      \\
      
      SYM1 & $C_2(\rho)$       &   0.1              &   $24.1\pm0.8$    & \cite{Trippa08}\\
      SYM2 & $C_2(\rho)$       & $0.01~\text{to}~0.19$       &   IAS,HIC         & \cite{Danielewicz14,Tsang09} \\
      SYM3 & $C_2(\rho)$       &   $0.01~\text{to}~0.31$      &   ASY-EoS       & \cite{Russotto16}\\
      
      \toprule
     \end{tabular}
\end{table}
In order to determine $\frac{m_0^*}{m}$ and $\Delta m_0^*$, one
needs to know the values of $k_+$ and $k_-$  [Eqs. (\ref{mstar2}) and
(\ref{dmstar1})].  The calculations are  performed in two stages. In
stage 1, the seven parameters $\tilde \al, K_1,K_2,a,b, k_+ $ and
$k_-$ occurring in the EDF given by Eq.(\ref{ener4}) are obtained from
optimization of the $\chi^2$-function from a fit to all the different
empirical and precision theoretical data listed in Table \ref{tab1}.
In the fitting protocol, in addition, values of  three empirical
nuclear constants pertaining to SNM (energy per nucleon $\epsilon_0$,
saturation density $\rho_0$ and incompressibility $K_0$) are further
chosen to be constrained; they are taken from the averages of the
'best-selected' nuclear EoS given in Ref. \cite{Dutra12}. Their values
are $\epsilon_0=-15.88 \pm 0.24$ MeV, $\rho_0=0.163 \pm 0.005$ fm$^{-3}$
and $K_0=226.2 \pm 10.1$ MeV.  These values refer to infinite symmetric
nuclear matter, but with roots embedded to finite nuclear observables.
Henceforth, these data would be referred to as 'macrodata'.  This
fitting protocol is seen to be incapable of yielding the values of
$k_+$ and $k_-$ uniquely, but gives a  value of a linear combination
of them (shown later). In stage 2, by fitting the 'constructed'
values of $m_1$ (see Eq. (\ref{ald2})) from chosen values of $E_x$
( discussed in subsection III B) for the considered nuclei, we get
$\Th_V$ which is a different linear combination of $k_+$ and $k_-$
as given by Eq. (\ref{tv1}). Combining  results from the two stages,
unique values of $k_+$ and $k_-$ are obtained.  The errors pertaining
to the studied observables are calculated from the curvature matrix
obtained from the double derivative of $\chi^2$-function with respect
to different parameters employing the method of covariance analysis
\cite{Dobaczewski14, Erler15,Mondal15}.

\subsection{Fitting of macrodata}

The macrodata (barring the ones at the saturation density)  
used in the fitting protocol are listed in 
Table \ref{tab1}.  The rows and 
columns are self explanatory. 
The first two rows refer to pressure of SNM. They are 
obtained from analysis of directed and elliptic flow \cite{Danielewicz02}
and kaon production \cite{Fuchs06,Fantina14} in heavy ion collisions
(HIC). 
The next four rows correspond to PNM. Its energy at a density
$\rho$ =0.1 fm$^{-3}$ is taken from the 'best-fit' Skyrme EDFs \cite{Brown13}.
The information on the energy and pressure of low density neutron matter
is taken from high precision predictions at next-to-next-to-next-to-
leading order ($N^3LO$) in chiral effective field theory \cite{Hebeler13,
Sammarruca15}. The pressure of PNM is the excess over the pressure of
SNM due to symmetry energy. It is constructed theoretically with two
extreme parameterizations, the soft (Asy Soft)  and the stiff
(Asy Stiff) symmetry energy \cite{Prakash88}. Its values are taken 
from Ref. \cite{Danielewicz02}. The last three rows refer to
the symmetry energy coefficients $C_2(\rho)$ at the densities
mentioned in the Table. They come from three different sources,
namely, simulation of low energy HIC in $^{112}$Sn+$^{112}$Sn and
$^{124}$Sn +$^{124}$Sn \cite{Tsang09,Tsang10}, nuclear structure
studies involving Isobaric Analogue States (IAS) \cite{Danielewicz14}
and Asy-EOS experiments at GSI \cite{Russotto16}. 
In addition, the
value of $C_2(\rho)$ at $\rho$ =0.1 fm$^{-3}$ quoted from microscopic
analysis of IVGDR in $^{208}$Pb is taken \cite{Trippa08} into consideration. 

\begin{figure}
\includegraphics[width=1.0\columnwidth,angle=0,clip=true]{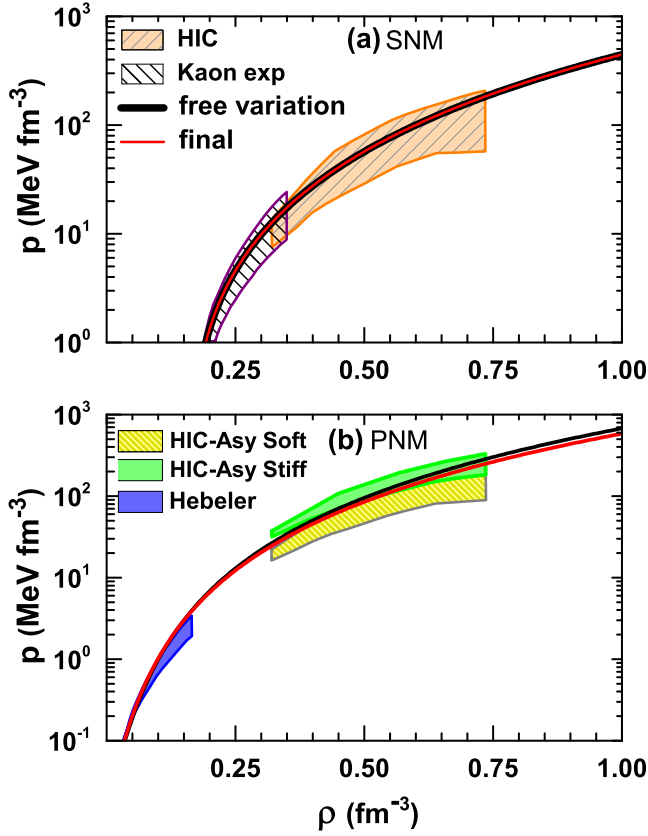}
\caption{\label{fig1} (color online) The pressure $P(\rho)$ for SNM (upper panel) 
and PNM (lower panel) as a function of baryon density $\rho$ for the 
best-fit parameters obtained from free variation of all the parameters 
(black lines) and for the final values of the parameters (see Table \ref{tab2}) 
shown by the red line. }
\end{figure}
\begin{figure}
\includegraphics[width=1.0\columnwidth,angle=0,clip=true]{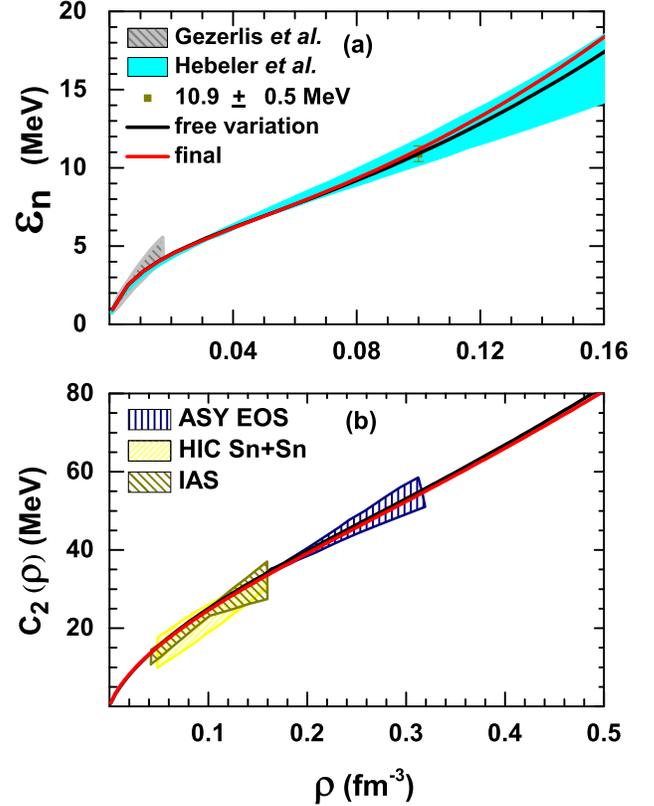}
\caption{\label{fig2} (color online) The energy per 
neutron $\epsilon_n$ of PNM (upper panel) and symmetry energy coefficient
$C_2(\rho)$ (lower panel) as a function of baryon density $\rho$. The
black and the red lines bear the same meaning as in Fig. \ref{fig1}.}
\end{figure}

A free variation of all the seven parameters with the above fitting
protocol yields a very shallow minimum in $\chi^2$ corresponding to
$\frac{m_0^*}{m} \simeq $ 1.31 and $\De m_0^* \simeq -2.9$ $\de $. The
fit to the empirical data is found to be very good as shown by black
lines in Figs. \ref{fig1} and \ref{fig2}.  The $\chi^2$-function is,
however, found to be very flat.  In order to get 
an insight into  this flatness
problem, we constrain $\tilde \alpha$ to a fixed
value and optimize $\chi^2$ varying the remaining six parameters.
This is similar to  the method adopted by Friedrich and Reinhard
\cite{Friedrich86} in finding out the interaction parameters of the
Skyrme EDF in their fitting protocol from their chosen data. They
found their routine  incapable of determining $\alpha $ ($=\tilde{
\alpha} -1$) and therefore had to be varied by hand to determine
the remaining Skyrme parameters.
We do likewise, we repeat the fitting calculations for different choices
of $\tilde \alpha $. Each choice of $\tilde \alpha $ leads 
to a different set of EDF parameters and thus $\frac{m_0^*}{m}$.
Each parameter set is found to be equally good in fitting the macrodata,
an unique value of $ \frac{m_0^*}{m}  $ can not thus be arrived at from
this fitting.  We find that $\frac{\De m_0^*}{\de}$ decreases with
increase in $\frac{m_0^*}{m}$.  The trend is found to be almost
parabolic in nature ( more on this is discussed later  in relation to
Fig.\ref{fig5}).

\begin{figure}
\includegraphics[width=1.0\columnwidth,angle=0,clip=true]{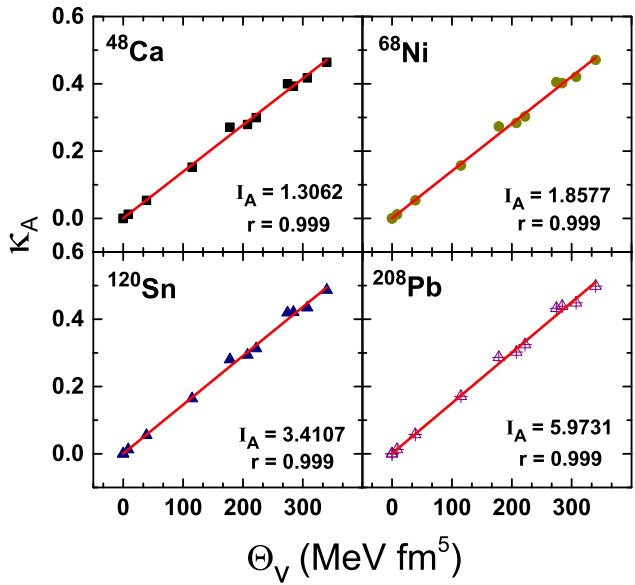}
\caption{\label{fig3} (color online) The correlation of the isovector
parameter $\Th_V$ obtained from the Skyrme EDFs \cite{Brown13}, with the
calculated dipole enhancement factor $\kappa_A$ for the nuclei $^{48}$Ca,
$^{68}$Ni, $^{120}$Sn and $^{208}$Pb. The corresponding values of the
integrals $I_A$ ( in units of fm$^{-3}$) and the correlation coefficients
are shown in  each panel.} \end{figure}

\subsection{Fitting of EWSR}

Eqs. (\ref{mp1}) and (\ref{kap1}) show that the isovector entity $\Th_V$
can be calculated if the EWSR sum $m_1$ and the integral $I_A $  are
known.  From the neutron and proton densities $\rho_n(r)$
and $\rho_p(r)$ calculated in the Hartree-Fock approximation for the four nuclei
{\it viz.} $^{48}$Ca \cite{Birkhan17}, $^{68}$Ni \cite{Rossi13},
$^{120}$Sn \cite{Hashimoto15} and $^{208}$Pb \cite{Tamii11} (for
which data on nuclear dipole polarizability are available) with the
'best-fit' Skyrme-EDF reported in Ref.\cite{Brown13}, it is found that
the integrals $I_A$ for a particular nucleus are nearly independent of
EDFs. This manifests in an extremely strong correlation  (with correlation
coefficient practically unity) between $\Th_V$ and $\kappa_A$ as displayed
in Fig. \ref{fig3}. The slopes of the correlation lines are taken as measures
for $I_A$ for each nucleus; they are shown in respective panels in
the figure.

\begin{figure}[h]
\includegraphics[width=1.0\columnwidth,angle=0,clip=true]{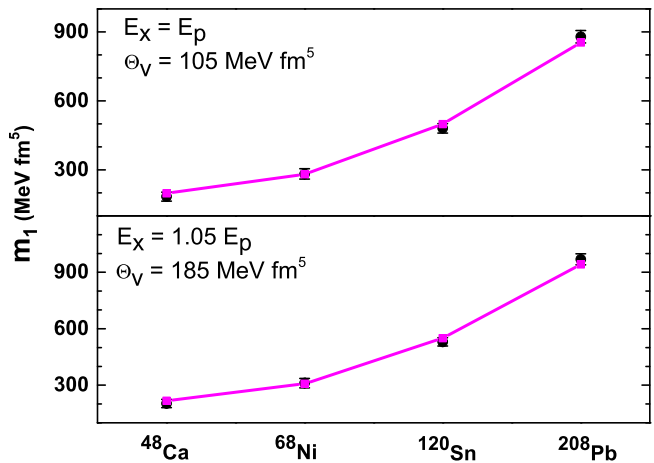}
\caption{\label{fig4} (color online)
 The calibrated values of EWSR
($m_1$) displayed for the four nuclei (black squares). The upper panel
shows the values with $E_x=E_p$, the lower panel displays the same with
$E_x=1.05E_p$ (see text for details). The solid lines are drawn to show
the fit with $\Th_V=105$ MeV fm$^5$ and 185 MeV fm$^5$, respectively.}
\end{figure}

Since the experimental values of $m_1$ are not very reliable due to the
contamination from 'quasi-deuteron effect', existing data on $\al_D$
for the four nuclei can be exploited to gauge the measures of $m_1$
in good bounds with reasonable choice of $E_x$. Two choices for its
values are made.  For its lower value, the known peak energy $E_p$ of the
experimental IVGDR strength function is chosen. For the higher value, we
take $E_x=1.05 E_p$. This choice is prompted  from our finding that RPA
calculations with the best-fit Skyrme EDFs \cite{Brown13} yield  $E_x$ to
be higher than $E_p$ by $\sim $ (4-6)$\% $ for the nuclei studied. These
two choices of $E_x$ give the lower and upper bounds of $m_1$.

First equating $E_x$ with $E_p$, the peak energy of the experimental
IVGDR strength distribution, $m_1$ are calculated for the four nuclei
from the experimentally obtained values of $\al_D$ (see Eq.(\ref{ald2}))
which are referred to as 'calibrated' values of $m_1$. Using these
calibrated values, the enhancement factor $\kappa_A$ for the chosen nuclei
are determined from Eq.(\ref{mp1}). With the known values of $I_A$ and
so obtained $\kappa_A$ are then subjected to a $\chi^2$ minimization
by varying $\Th_V$ (Eq.(\ref{kap1})). The optimized value of $\Th_V$
is found to be $\Th_V =$105.0 MeV fm$^5$.  The calculation is repeated
with $E_x$ increased by 5 $\%$ above the values of $E_p$. The optimized
value of $\Theta_V$ is now 185.0 MeV fm$^5$.  The fitted results with
the two sets of calibrated values of $m_1$ are shown in the upper and
lower panels of Fig. \ref{fig4}. In both cases the fits are very good.
An average value of $\Theta_V \simeq 145.0 \pm 40.0 $ MeV fm$^5$ can be
inferred from the calculations.  Since $\Th_V$ determines the difference
between $k_+$ and $k_-$,  its constancy demands that if $k_+$ increases,
$k_-$ should also increase.

\begin{figure}
\includegraphics[width=1.0\columnwidth,angle=0,clip=true]{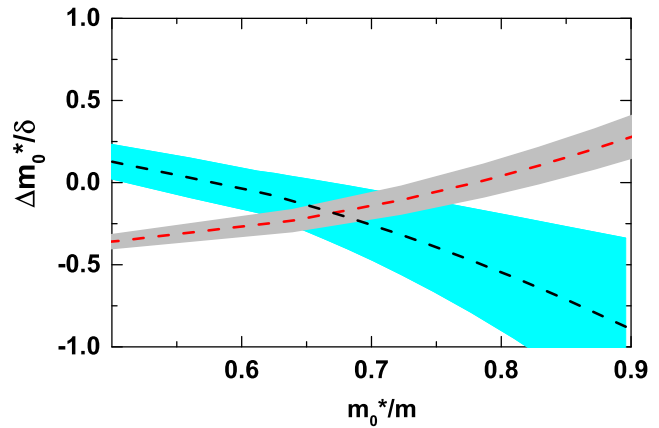}
\caption{\label{fig5} (color online) The isovector effective
mass-splitting as a function of nucleon effective mass. The black dashed
line refers to the best fit obtained from macrodata for different values
of $\tilde \alpha$; the red dashed line corresponds to the one obtained
by satisfying Eq. (\ref{dmstar3}) with $\Th_V = 145$ MeV fm$^5$. The
cyan and grey shades refer to the respective uncertainties. }
\end{figure}

Using Eq.~(\ref{mstar2}), (\ref{dmstar1}) and (\ref{tv1}), one gets
\bea
\label{dmstar3}
\frac{\De m_0^*}{\de}= \left(\frac{m}{\hbar^2}\Th_V \rho_0 -2\frac{m}
{m_0^*} +2 \right) \left(\frac{m_0^*}{m}\right)^2.
\eea
One finds increase in $\frac{m_0^*}{m}$ with increase in 
$\frac{\De m_0^*}{\de}.$  This is complementary to what we obtained from fitting
the macro data. As  mentioned earlier, there we find that $\frac{\De
m_0^*}{\de}$ decreases with increasing  $\frac{m_0^*}{m}$ almost in a
parabolic way, it can be  well approximated as,
\bea
\label{dmstar2}
\frac{\De m_0^*}{\de} = \beta_1 +\beta_2 \left(\frac{m_0^*}{m}\right)^2,
\eea
 with $\beta_1 =0.733 \pm 0.024 $ and $\beta_2 =-2.029 \pm 0.032$.
This equation can be restated as 
\bea
\label{b1b2}
(k_-+\beta_1 k_+)\rho_0 \simeq -(\beta_1+\beta_2). 
\eea
Since $\beta_1$ and the r.h.s.  of this equation are positive,
one finds that if $k_-$ increases, $k_+$ decreases and vice versa.
The opposing trends on the relation  of $\frac{\De m_0^*}{\de}$ on
$\frac{m_0^*}{m}$ from Eqs. (\ref{dmstar2}) and (\ref{dmstar3}) are
displayed in Fig.\ref{fig5}.  The black dashed line corresponds
to Eq. (\ref{dmstar2}), the red dashed line corresponds to
Eq. (\ref{dmstar3}) with $\Th_V=145.0$ MeV fm$^5$. The lower and
upper boundaries of the grey shade around the red dashed line refer to
calculations with $\Th_V=105.0$ and 185.0 MeV fm$^5$, respectively. The
cyan shade around the black dashed line corresponds to uncertainties
involved in relation to parameter fitting.

The intersection of the dashed black and red lines yields the central
values of both $\frac{m_0^*}{m}$ and $\frac{\De m_0^*}{\de}$; $k_+$
and $k_-$ are then known. With the constraints on $\epsilon_0,~\rho_0$
and $K_0$ as mentioned earlier and with known $k_+$ and $k_-$, the other
parameters of the EDF are then determined from the optimization of the
$\chi^2$-fit to the macrodata given in Table \ref{tab1}.

\begin{table}
 \caption{The final model parameters obtained by optimizing the
$\chi^2$ function together with the uncorrelated and correlated errors
(see text for details). The parameters  $K_1$ and $K_2$ are in units
of MeV fm$^{3}$, $a$ and  $b$ are in MeV fm$^{3\tilde{\alpha}}$ and $k_+$ and
$k_-$ are in fm$^{3}$.}
  \label{tab2}
\setlength{\tabcolsep}{2.2pt}
\renewcommand{\arraystretch}{1.1}
\begin{tabular}{ccccccc}
\toprule
	    
 ${\tilde \alpha}$ & $K_1$ & $K_2$ & $a$ & $b$ & $k_+$   &$k_-$  \\
	  \hline
	    
  1.11 & -1220.21 & 977.94 & 120.03 &  -121.93 & 6.07  & 2.60  \\
Unc. err. & 1.16     & 2.38       & 0.15          &  0.33         &   0.10
&    0.15       \\
          Cor. err. &  103.04    &  90.25       &  15.01     &  13.57
&  1.13       &  0.96      \\

\toprule 
 \end{tabular}
\end{table}

\begin{figure}[h]
\includegraphics[width=1.1\columnwidth,angle=0,clip=true]{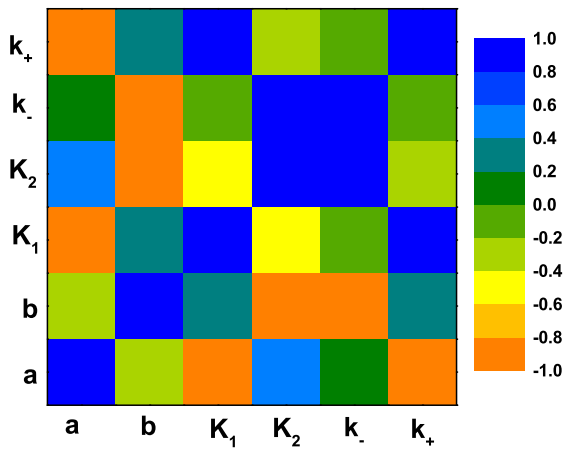}
\caption{\label{fig6} The correlation among various model  parameters. The
values of the correlation coefficients are colour coded. }
\end{figure}

\begin{table}
   \caption{Different properties pertaining to nuclear matter (NM) and 
            neutron star (NS) obtained with the final parameters listed in Tab. \ref{tab2}.}
   \label{tab3}
    \setlength{\tabcolsep}{12.5pt}
      \renewcommand{\arraystretch}{1.1}
    \begin{tabular}{lccc}

   \toprule
 Type   & & Unit &Value \\
        \hline
  NM &  $\epsilon_0$   & MeV                    & $-15.93\pm0.20$       \\
     &  $\rho_0$ & fm$^{-3}$            & $0.1620\pm0.003$      \\
     &  $K_0$   & MeV                   & $225.23\pm6.38$       \\
     &  $m_0^*/m$ &                     & $0.68\pm0.04$ \\
     &  ${m_{v,0}^*}/{m}$ &         & $0.78 \pm^{0.05}_{0.04}$\\
     &  $\Delta m_0^*/ \delta$ &        & $-0.20\pm0.09$        \\
     &  $C_2(\rho_{0})$  & MeV          & $33.94\pm0.50$        \\
     &  $L_0$   & MeV                   & $68.50\pm3.72$        \\
     &  $K_{\rm sym}^0$ & MeV           & $-47.46\pm17.87$      \\
     &  $K_\tau$ & MeV                  & $-349.22\pm14.06$     \\
     &  $M_{\rm c}$   & MeV             & $998.79\pm41.29$      \\
     &  $Q_0$   & MeV                   & $-359.23\pm23.08$     \\

     \hline
  NS  & $M_{\rm max}^{\rm NS}$ & $M_\odot$      & $2.06\pm0.03$ \\
      & $R_{1.4}$ & km  & $12.62\pm0.57$          \\
        \toprule

  \end{tabular}
  \end{table}

From the crossing of the shades as shown in Fig. \ref{fig5}, the values
of $\frac{m_0^*}{m}$ and $\frac{\De m_0^*}{\de} $ are found to be in the
range $0.61$ to $0.75$ and $ -0.3$ to $-0.1$, respectively.  The final
values of the  parameters corresponding to the projected central values
of $\frac{m_0^*}{m}$ and $\frac{\De m_0^*}{\de}$ from Fig.~\ref{fig5} are
listed in Table \ref{tab2}.  The value of $\tilde \alpha $ comes out to
be 1.11.  The uncorrelated and correlated errors of the parameters
obtained within the covariance method are also presented.  We see that
the correlated errors are significantly higher in comparison to the
uncorrelated ones which indicate the existence of strong correlations
among the parameters. In Fig. \ref{fig6}, the correlation among the model
parameters are depicted in terms of the Pearson's correlation coefficient.
Two parameters  are said to be fully correlated if the magnitude of the
correlation coefficient for them is unity as shown by the orange and blue
colours. The parameters $a, K_1$ and $k_+$ are strongly correlated among
themselves; the same is true for the parameters $b, K_2$ and $k_-$.
With the parameters presented in Table \ref{tab2} the obtained fit
to various macrodata  are displayed by red lines in Figs. \ref{fig1}
and \ref{fig2}. One can see that the difference between the fits to the
data from free variation (black line) and the calculation with the final
parameters (red line) is insignificant.  The isovector mass comes out
to be $\frac{m_{v,0}^*}{m}=0.78  \pm ^{0.05}_{0.04}$.

With the parameters of the EDF as listed in Table \ref{tab2}, the
values of symmetry energy coefficient $C_2(\rho_0)$, its density slope
$L_0\left(=3\rho \frac{\partial C_2}{\partial \rho}|_{\rho_0}\right)$,
the curvature parameter $K_{sym}^0\left(=9\rho^2\frac{\partial^2
C_2}{\partial \rho^2}|_{\rho_0}\right)$ and the symmetry
incompressibility at saturation corresponding to asymmetric nuclear matter
$K_\tau(=K_{sym}^0-6L_0-Q_0L_0/K_0)$ are calculated. Here $Q_0 \left(=27
\rho_0^3{\frac{\partial^3\epsilon}{\partial \rho^3}} |_{\rho_0}\right)$
is the skewness parameter corresponding to SNM.  Their values are listed
in Table \ref{tab3}.  All of them are seen  to lie within the accepted
range obtained from different EDFs \cite{Dutra12}. The value of the
derivative of incompressibility $M_c\left(=3\rho \frac{dK}{d\rho }
|_{\rho_c}\right)$ for SNM at a sub-saturation density $\rho_c \simeq
(0.710\pm 0.005)\rho_0$ also has excellent agreement with that obtained
from examination of isoscalar giant monopole resonance (ISGMR) data
for $^{208}$Pb and $^{120}$Sn \cite{Khan12, Khan13}. For completeness,
to gauge the applicability of the EDF at extremely high density, the
maximum mass of the neutron star ($M_{max}^{NS}$) is also calculated.
The EOS of the crust is taken from the Baym, Pethick and Sutherland model
\cite{Baym71}. The EOS of the core is calculated with the assumption
of a charge-neutral uniform plasma of neutrons, protons, electrons and
muons in $\beta-$equilibrium. The value of $M_{max}^{NS}$ is seen to
be $(2.06\pm 0.03)M_\odot$, in consonance with the recently observed
maximum neutron star mass \cite{Demorest10,Antoniadis13}. The value of
the radius $R_{1.4}$ of a neutron star of mass $1.4 M_\odot $ is also
in tune with the constrained value obtained from analysis of different
models \cite{Lattimer13}. In passing, it is mentioned that recently
a new Skyrme EDF is proposed \cite{Zhang18}. It is commensurate with
predictions from chiral effective field theory, binding properties of
finite nuclei and also the electric dipole polarizability.  The effective
mass is $\frac{m_0^*}{m}= 0.75 \pm 0.04$; the isovector splitting of
the  effective mass is positive, $\sim 0.12 \delta$. However, we find
it to be incompatible with the criterion for the observed maximum mass
of the neutron star. The mass turns out to be $1.8 M_\odot$, some what
below the experimentally observed maximum mass.
\section{summary and conclusions}

We have proposed a means of finding out the value of the nucleon
effective mass $m_0^*$ and its isovector splitting $\De m_0^*$ by
using a form of EDF \cite{Mondal17} built without any reference to
any particular interaction  but with a few plausible assumptions on
its nature.  The structure of the EDF is seen to be equivalent to the
'standard' Skyrme functional under certain approximations.  We work in
the framework of this energy functional and find its parameters  from
$\chi^2$- minimization of the empirical nuclear matter data and the
existing `state of the art' theoretical data pertaining to neutron matter.
It is observed that the fit to these data is unable to determine $m_0^*$
and $\De m_0^*$ unambiguously, but yields  a well-tuned combination of
them; an almost indiscernible fit to the macrodata   can be obtained
over a wide range of their values.

From experimental data related to nuclear dipole polarizability, we
show how this veil of indeterminacy can be lifted. These particular
data on finite nuclei, if used judiciously give information on a linear
combination of parameters determining the nucleon effective mass and
its isovector splitting that is complementary to what was obtained in
relation to the macrodata and thus can project out the values of the
nucleon effective mass and its isovector splitting within reasonable
constraints. In doing so, there is no compromise in the excellent
agreement of the predicted values of the nuclear constants related to
symmetric and asymmetric nuclear matter with the ones broadly accepted
in present day wisdom, nor there is any sacrifice in the fit to the
empirical data related to neutron stars.

\section{Acknowledgments} T.M. is grateful to the Saha Institute of
Nuclear Physics for the hospitality accorded to him during the phase of
this work.  J.N.D. acknowledges support from the Department of Science
and Technology, Government of India  with grant no. EMR/2016/001512.


\end{document}